\documentclass[11pt]{article}
\pdfoutput=1

\usepackage{xcolor}
\usepackage{textcomp}
\usepackage{cite}
\usepackage{multirow}
\usepackage{enumerate}
\usepackage[margin=2.75cm]{geometry}

\usepackage{graphicx}
\usepackage[subrefformat=parens,labelformat=parens]{subfig}

\usepackage{enumitem} 

\usepackage{amsfonts,amsmath}
\usepackage{multirow,multicol}
\usepackage{bbm}
\usepackage{xspace}
\usepackage{slashed}

\usepackage{float}
\usepackage{array}
\newcolumntype{P}[1]{>{\centering\arraybackslash}p{#1}}
\newcolumntype{M}[1]{>{\arraybackslash}m{#1}}

\usepackage[colorlinks = true, linkcolor = blue, urlcolor = blue, citecolor = blue, anchorcolor = blue]{hyperref}

\begin{document}
\begin{center}
{\Large\bfseries Evolving neural networks with genetic algorithms to study the String Landscape}\\[7mm]
\textsc{Fabian Ruehle\,\footnote{fabian.ruehle@physics.ox.ac.uk}}\\[3mm]
\textit{Rudolf Peierls Centre for Theoretical Physics, Oxford University,\\ 1 Keble Road, Oxford, OX1 3NP, UK}
\end{center}

\begin{abstract}
\noindent We study possible applications of artificial neural networks to examine the string landscape. Since the field of application is rather versatile, we propose to dynamically evolve these networks via genetic algorithms. This means that we start from basic building blocks and combine them such that the neural network performs best for the application we are interested in. We study three areas in which neural networks can be applied: to classify models according to a fixed set of (physically) appealing features, to find a concrete realization for a computation for which the precise algorithm is known in principle but very tedious to actually implement, and to predict or approximate the outcome of some involved mathematical computation which performs too inefficient to apply it, e.g.\ in model scans within the string landscape. We present simple examples that arise in string phenomenology for all three types of problems and discuss how they can be addressed by evolving neural networks from genetic algorithms.
\end{abstract}

\section{Introduction}
\label{sec:Introduction}
Despite the fact that we have created over the years an enormous set of data in string theory, be it construction mechanisms and data bases for huge classes of Calabi-Yau manifolds \cite{Candelas:1987kf,Kreuzer:2000xy,Morrison:2012js,Morrison:2012np,Gray:2013mja,Anderson:2015iia} or string models and flux vacua \cite{Davey:2009bp,Anderson:2013xka,Nilles:2014owa,Nibbelink:2015ixa,Blaszczyk:2015zta,Franco:2016qxh,Halverson:2016tve,Halverson:2017ffz} in the string landscape \cite{Douglas:2003um,Kachru:2003aw}, we have until now \cite{He:2017aed,Krefl:2017yox} not applied machine learning techniques to analyze this data. As exemplified by the large amount of work that has gone into constructing MSSM-like models from string theory, the problem is that string theory seems to lack a selection mechanism that can be used to choose among its many vacua. 

If we ever want to be able to test string theory, it is crucial to know which \textit{generic} predictions are common among all models derived from it (such as extra dimensions, hidden sector gauge groups, additional fields). Once we have identified such generic features we can try to devise methods for either testing these features directly or study their implications and consequences in order to find common properties that are shared by many string vacua. Thus, a long-term goal would be to use machine learning to actually analyze and search for common patterns in all these constructions. As the number of models and the details we know about them grow, it is conceivable to feed the physical properties of the models into a machine learning algorithm and let it analyze common features. Alternatively, one could use machine learning to produce new models with good features from known models. While this might be hard in some cases (e.g.\ using a given F-theory model on a Calabi-Yau fourfold with $G_4$-flux to predict a good candidate for an MSSM-like model that is constructed via a non-geometric CFT description of string theory), others might be more accessible, especially when we stay within the same description of string theory. However, there are also some questions that need to be addressed in many formulations of string theory, such as knowledge of the topological data defining the underlying compactification space. 

In order to address such questions, we study artificial neural networks (ANNs), which feature prominently in machine learning. ANNs were designed to mimic the way the human brain works. An ANN consists of several layers. Each layer receives some input, performs a mathematical operation on it to produce some output which is passed on to the next layer. The precise way in which these ANNs learn and function is a current active field of research. ANNs can be used in different ways:
\begin{enumerate}[label=(\Alph*)]
\item Classify input data. This means that the ANN knows a finite set of classes into which the input is grouped. \label{enum:classify}
\item Study relations between input and output. In some cases, we know in principle how some input data (e.g.\ flux) is related to the output data (e.g.\ the string spectrum), but the concrete computation is very involved. \label{enum:reverse-engineer}
\item Bypass very slow or unfeasible steps in model scans by predicting the outcome of this computation (such as e.g.\ computing bundle cohomologies) based on regression. \label{enum:predict}
\end{enumerate}

The reason why ANNs can be applied to these questions is due to the universal approximation theorem \cite{Cybenko:1989aa}, which states that a sufficiently complex ANN can approximate any function to an (in principle) arbitrarily high precision. The advantage of using trainable ANNs over common interpolation methods is that one does not have to decide on the details of the interpolation function, or even which of the many features of the input data one should actually interpolate.

Despite the fact that ANNs can in principle be applied to all three questions, it is a priori not clear what the neural network should look like (how many layers, which layer sizes, which types of layers, \ldots) to perform best for the specific task it is given. For example, depending on whether we want to apply the ANN to a classification or regression problem, some layers might be more useful than others. When classifying data, i.e.\ predicting the most likely outcome from a fixed set of possible outcomes, the neural network can be set up to map the input data to a vector containing the probability for each of the possible outcomes and the one with the largest probability is chosen (this is sometimes referred to as \textit{one-hot encoding}). In contrast, when trying to predict results from an infinite or continuous set (such as $\mathbbm{Z}$ or $\mathbbm{R}$), the ANN has to either reproduce the exact computational steps that maps the input to the output or it has to interpolate the input to produce a viable output, which is possible at least in principle by virtue of the universal approximation theorem \cite{Cybenko:1989aa} (we call ANNs of this type \textit{predictors}). While the set of operations might be known in some cases, it is unknown in many applications. This begs the question of how we should ever find a good neural network if we do not know which building blocks or network topology to use for the best possible outcome.

The idea is to copy nature yet again and to use genetic algorithms to \textit{evolve} a neural network that performs best for the task it was given. The idea of genetic algorithms is to start with an initial generation that evolves through reproduction and mutation into a population that is fittest with respect to a given task. So far, genetic algorithms have, just as ANNs, been barely applied by string theorists, with the notable exceptions of \cite{Allanach:2004my,Abel:2014xta}. 

A huge advantage of ANNs is that each node is independent from the other nodes in a given layer, which makes parallelization of the computation over layers straight-forward. Also note that many existing libraries for ANNs (including the one for \texttt{Mathematica} we are using here) come with GPU support, which can further speed up the computations. Likewise, the implementation of the genetic algorithms can be parallelized over the individuals within each generation.

The remainder of this paper is organized as follows. In Section \ref{sec:ANN} we provide a basic introduction to ANNs, explain our conventions and present some toy examples to introduce the ideas and illustrate how ANNs can be used in applications of type \ref{enum:classify} and \ref{enum:reverse-engineer}. These examples are chosen to be rather simple and pedagogical in order to illustrate the key steps. In Section \ref{sec:GeneticAlgorithms} we introduce genetic algorithms and explain how we apply them to ``breed'' effective neural networks. We then utilize these techniques in an example in Section \ref{sec:Example}, where we look at the evolution of an ANN that can predict line bundle cohomology based on regression.

\section{Artificial neural networks}
\label{sec:ANN}

\begin{figure}[t]
\centering
  \subfloat[][Simple feed-forward ANN with 6 layers. Top: Connections and nodes drawn explicitly. Bottom: Same ANN in the layer representation.]{\shortstack{\includegraphics[width=.54\textwidth]{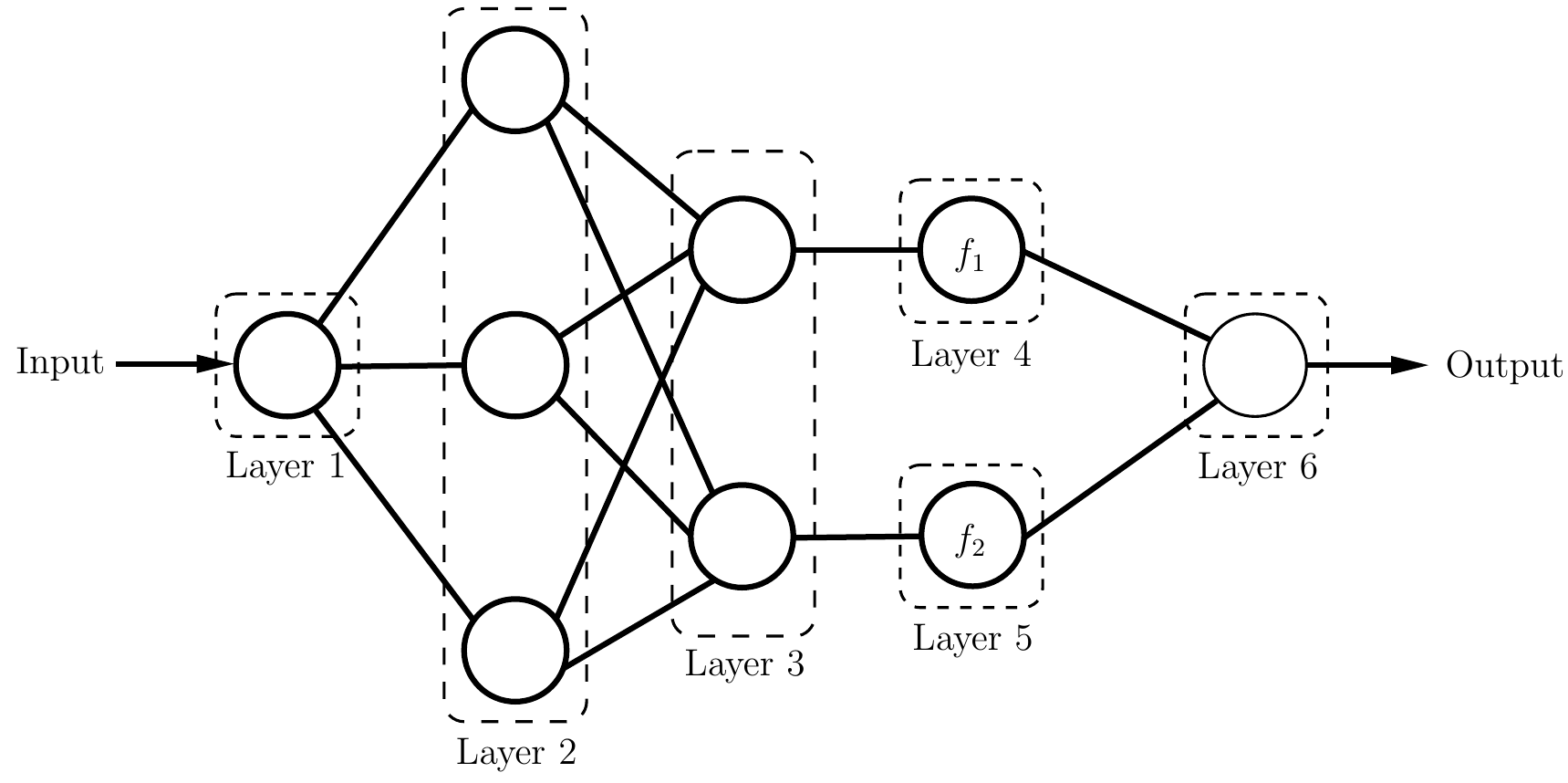}\\ \includegraphics[width=.54\textwidth]{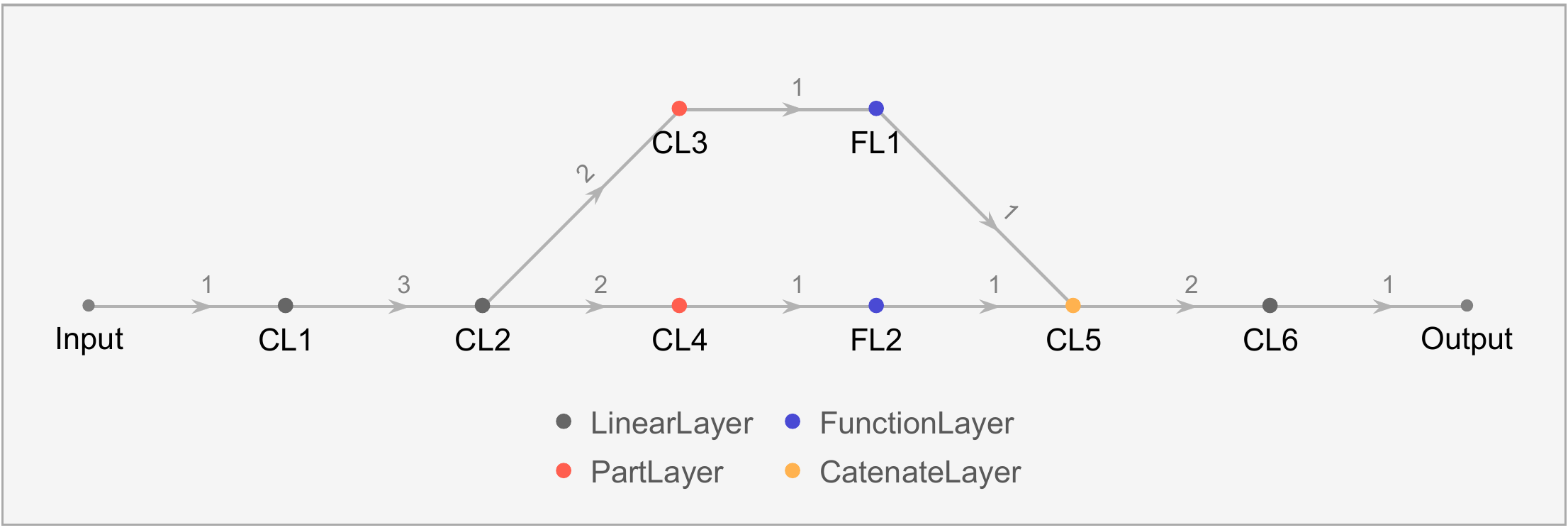}}\label{fig:FFNNGeneric}}\qquad\quad
  \subfloat[][Simple example of a recurrent neural network featuring a feed-back loop.]{\shortstack{\includegraphics[width=.35\textwidth]{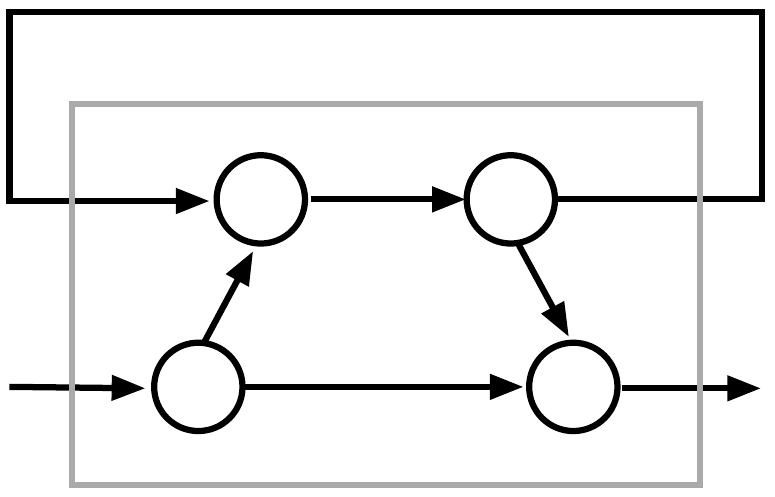}\\[10mm]~}\label{fig:RNNGeneric}}
  \caption{Two example ANN implementations: \protect\subref{fig:FFNNGeneric} shows a feed-forward ANN with connections between nodes given explicitly (top) and the same graph in the layer representation (bottom). \protect\subref{fig:RNNGeneric} shows a simple example of a recurrent ANN.}
  \label{fig:ANNGeneric}
\end{figure}

In this Section, we review basics of artificial neural networks, introduce our notation and present simple examples. ANNs can be thought of as a function that maps some given input to some output. In their simplest form, the input and output are just vectors of data, e.g.\ 
\begin{align*}
\text{ANN}: \mathbbm{Z}^n\rightarrow\mathbbm{Z}^m\,.
\end{align*}
The ANN is organized in layers. The first layer is called the \textit{input layer} and the last layer is called to \textit{output layer}. All layers that occur in between are \textit{hidden layers}. The various layers are connected via ports. In linear (feed forward) ANNs, the output port of one layer is connected to the input port of the next layer. More complicated scenarios are of course possible as well. For example, the ANN could split its input at some hidden layer into $n$ different branches, perform an independent number and type of operation on each branch separately, and join the various branches at some other hidden layer, see e.g.\ Figure~\ref{fig:ANNGeneric}\subref{fig:FFNNGeneric}. Another extension of this idea would be to create loops, i.e.\ to split the input at some layer, perform some operation on the two branches (that possibly mixes them partially) and feeds part of the result back into the original layer (this is called feed-back). Such neural networks are also referred to as recurrent neural networks\footnote{We will always be using the term ANN even if our artificial neural network contains recurrent layers.} (RNNs), see e.g.\ Figure~\ref{fig:ANNGeneric}\subref{fig:RNNGeneric}. A prominent example of an RNN is a Long Short-Term Memory Layer (LSTM). 

A \textit{general layer} consists of a collection of nodes or \textit{perceptrons}. Mathematically, the different layers are connected via a linear transformation,
\begin{align}
v_{\text{output}}=A\cdot v_{\text{input}}+b\,
\label{eq:LayerLinTrafo}
\end{align}
i.e.\ matrix multiplication by $A$ and addition of a constant vector $b$, the so-called \textit{bias}. Each layer (or more precisely each perceptron in each layer) applies a function, the so-called \textit{activation function}, to the output of this linear transformation. 

In the following, especially for the combination of ANNs with genetic algorithms, we will heavily exploit the modular architecture of ANNs, i.e.\ their implementation in terms of layers that apply some (a priori arbitrary) function. A simple feed-forward ANN consists of a collection of generic layers where all nodes in the same layer apply the same activation function and all nodes of one layer are connected with all nodes of the next layer. To fully exploit the modular architecture, it proves very beneficial to also think of the connection between generic layers themselves as a special type of layer. This leads to what we call the \textit{layer representation} of an ANN where both its connections between layers and the activation function of its nodes are given in terms of layers. We adopt the convention that all layers are always fully connected to the next layer and a layer always applies a certain function to its entire input vector. Depending on the type of layer, it applies its function to all components simultaneously or to each component of the vector separately.

We will call layers that apply a function (such as the activation function of a generic layer) to their input \textit{function layers} and layers that describe the connections of the graph \textit{connection layers}.  In this way, we have replaced the connection of a layer to its predecessor and the activation function by two consecutive layers: a connection layer that encodes the connection and a function layer that applies the activation function. The most important connection layer is the \textit{linear layer}, which describes the usual connection between layers via a linear transformation of the type \eqref{eq:LayerLinTrafo} (note that this layer could also be thought of as a function layer, since it applies a linear function to all of its input values; however, we choose to focus on its role of connecting different layers and classify it as a connection layer). In the terminology above, a linear layer would correspond to a general layer where the activation function is trivial (i.e.\ the identity). Similarly, a function layer corresponds to a general layer where the linear transformation \eqref{eq:LayerLinTrafo} corresponding to the connection of this layer to the previous is fixed to be the identity, i.e.\ $A=\mathbbm{1}$ and $b=0$.

The layer representation allows us to represent the network corresponding to the ANN in a simpler way. For example, a simple deep feed-forward ANN that consists of several general layers for which all perceptrons in each layer apply the \textit{same} activation function will be simply given in terms of a chain of layers, consisting of alternating linear layers and function layers. Simple topological information of the ANN is also encoded in the linear layer: if e.g.\ the output of a node $a$ from layer $l_i$ is not connected to the input of node $b$ of layer $l_{i+1}$, the matrix corresponding to the linear layer $l+1$ has a zero at position $(b,a)$. However, in order to describe ANNs with a more complex topology, or ANNs in which different nodes of the same layer apply \textit{different} activation functions, we need to introduce more connection layers beyond linear layers. In contrast to the linear layer, these do not (necessarily) have free parameters (such as the entries of the matrix $A$ and vector $b$ of the linear layer). For example, we introduce \textit{part layers} that can split their input into $k$ disconnected parts, \textit{catenate layers} that join their inputs, \textit{reshape layers} that can transform the input vector into a tensor, etc.

Note that if several linear layers $l_1,\ldots,l_n$ occur consecutively, they can be combined into a single linear layer $\widetilde{l}$ whose linear function is given by the concatenation of the linear functions of the original layers,
\begin{align}
\label{eq:LinLayerCombi}
\widetilde{l}=l_n\circ\ldots\circ l_2 \circ l_1\,,\quad v_{\text{output}}=A_n\cdot(\ldots(A_2\cdot(A_1\cdot v_{\text{input}}+b_1)+b_2)\ldots)+b_n\,.
\end{align}
Similarly, if several function layers occur consecutively they can be combined into a single layer that simply concatenates these functions,
\begin{align}
v_{\text{output}}=f_n(\ldots(f_2(f_1(v_{\text{input}}))\ldots)\,.
\end{align}
In typical cases, neither $A$ nor $f$ are (fixed to be) trivial.

Combining layers along the lines discussed above, or used for example when representing the RNN in terms of a simple block in Figure \ref{fig:ANNGeneric}\subref{fig:RNNGeneric}, already foreshadows the much more general concept we will apply when combining ANNs with genetic algorithms. In principle, we can combine an arbitrary set of (connection and function) layers and use them as a new building block in a more complex ANN. Taken to the extreme, we could combine the entire ANN into just one building block and represent the ANN by it. While this is of course of little use we find it very beneficial to combine or group several (connection and function) layers that form a specific functional unit into one building block. For example, if we wanted to build an ANN that computes the factorial $k!$ of its input $k$, we might build it from two types of higher level building blocks; one which takes care of multiplying two components and a second that takes care of subtracting one from its input component. Another example for a layer that consists of several connection and function layers are RNN layers like the LSTM layer.

Once we have built our ANN we need to train it. During training, we provide a set of input and output values which the neural network can use to learn their correlation. Depending on the type of layer, i.e.\ on the function this layer applies to its input data, a layer might or might not have trainable parameters. For example, the parameters of the linear layers, i.e.\ the entries of the matrix and the bias vector, are typically taken to be trainable parameters while other connection layers (such as the part layer) have no trainable parameters. Of course, we can introduce a variety of other layers with trainable parameters, e.g.\ \textit{constant array layers} which do not have any input but output a constant but trainable array of real numbers. After each round of training the ANN updates the trainable parameters in order to minimize the error (e.g.\ by minimizing the mean squared distance between the training input and output). This is how the neural network learns. 

\bigskip
\noindent\textbf{Example: A simple feed-forward ANN}\\\noindent 
To give an example we look at the feed-forward ANN in Figure \ref{fig:ANNGeneric}\subref{fig:FFNNGeneric}. Let us discuss the representation on the top first. The input in this example is just a single number. By multiplying the input with an $N_1\times1$ matrix (and possibly adding a constant bias) this is fed into layer 2. Its output is thus a vector of length $N_1$ (in our example $N_1=3$). In a general layer, each node of layer 2 now applies an activation function, which we assume to be trivial for the sake of this example. The connection with the next layer corresponds to a multiplication with an $N_2 \times N_1$ matrix and addition of a constant bias, thus producing a vector of length $N_2$ (in our example $N_2=2$). After that, each node applies again its activation function. If we assume this activation function to be trivial as well, we could combine layers 2 and 3 into one layer as explained in \eqref{eq:LinLayerCombi}. In the transition to the next layer(s) the output of layer 3 is split into two pieces of length $N_2/2$ each and we branch the ANN. The corresponding matrix would be block diagonal with blocks of size $N_2/2\times N_2/2$ each. In a general layer, we could combine layers 4 and 5 into one layer whose first $N_2/2$ nodes apply the activation function $f_1$ and whose second $N_2/2$ nodes apply the activation function $f_2$. In the last step, the output of layers 4 and 5 are again mapped onto layer 6 via a simple linear transformation corresponding to a $1\times N_2$ matrix (again plus possibly a constant bias), which maps the vector back onto a single number. If we take the activation function of this layer to be again trivial this layer gives the output of our ANN. 

Next, we discuss the same ANN in its layer representation as given at the bottom of Figure~\ref{fig:ANNGeneric}\subref{fig:FFNNGeneric}. The layer labeled ``Input'' in the bottom graph corresponds to layer 1 of the top graph. The next layer CL1 is a linear layer of size 3 and corresponds to the connection of layers 1 and 2 in the top graph. Since we have assumed in this example that layer 2 of the top graph applies the identity activation function we need not include a function layer in the bottom graph that corresponds to this trivial activation function. The next layer in the bottom graph, CL2, is again a linear layer (this time of size 2), that describes the connection between layers 2 and 3 in the top graph. Again, we assumed a trivial activation function and thus need not include a corresponding function layer. At this point, the top graph splits, i.e.\ not all nodes of layer 3 are connected with all nodes of the next layer(s). To describe this in the layer representation of the bottom graph we include two part layers (labelled CL3 and CL4). By convention, both are supplied with the full output of layer CL2. To arrange the splitting, layer CL3 selects the first component while CL4 selects the second. This is then passed on to layers FL1 and FL2, which correspond to layers 4 and 5 in the top graph. FL1 and FL2 are function layers that apply functions $f_1$ and $f_2$, respectively. In the next step in the top graph, layers 4 and 5 are connected with layer 6. In the layer representation at the bottom, we first include a connection layer CL5 that joins the output of layers FL1 and FL2 and then feed this into the linear layer CL8, which corresponds to the connection of layers 4 and 5 with 6 in the top graph.

\begin{figure}[t]
\centering
  \subfloat[][All nodes and connections in the ANN.]{\qquad\includegraphics[width=.33\textwidth]{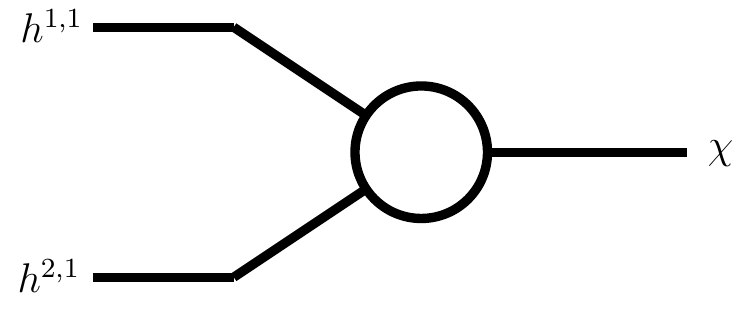}\label{fig:ANNEuler1}\qquad}\quad\qquad\qquad
  \subfloat[][Layer representation of the ANN.]{\includegraphics[width=.35\textwidth]{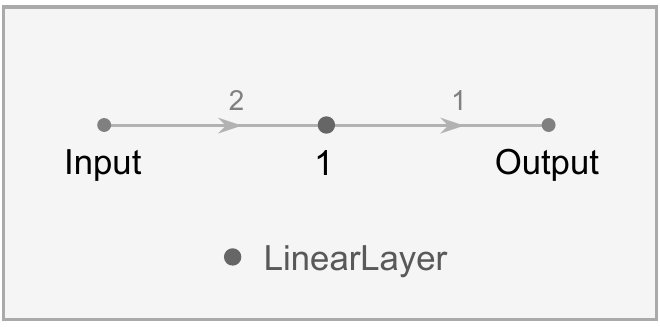}\label{fig:ANNEuler2}}
  \caption{Layout of the ANN that computes the Euler number $\chi$ from the Hodge numbers $h^{1,1}$ and $h^{2,1}$. In \protect\subref{fig:ANNEuler1} we show the full ANN with all nodes and connections. In \protect\subref{fig:ANNEuler2} we show the same ANN using a linear layer.}
  \label{fig:ANNEuler}
\end{figure}

\bigskip
\noindent\textbf{Example: Predicting the Euler number from $\boldsymbol{h^{1,1}}$ and $\boldsymbol{h^{2,1}}$}\\\noindent 
In order to see an ANN at work let us look at a simple example, where an ANN learns to compute the Euler number $\chi$ of a Calabi-Yau threefold from $h^{1,1}$ and $h^{2,1}$, i.e.\ $\chi=2(h^{1,1}-h^{2,1})$. This means that the input vector consists of two non-negative integers and the output vector is an integer. This simple example is equivalent to solving the corresponding linear regression problem. The transformation between the input and output vector is given by the matrix $A=(2,-2)$. We set up the neural network with one layer with two nodes and connect all input ports to all output ports, which leads to the ANN depicted in Figure~\ref{fig:ANNEuler}. The connection of the input to the first (and only) layer corresponds to a $2\times1$ matrix $A_1$ (to keep this introductory example as simple as possible, we set the bias vector $b$ to zero and make it non-trainable). This layer applies an activation function which we choose to be trivial for the sake of simplicity. Thus, the ANN provides a map from $\mathbbm{Z}^2$ to $\mathbbm{Z}$. In layer representation, this ANN consists of just a single linear layer of size 2.

The matrix $A_1$ is initialized with random real values. During training the ANN updates these values to get closer to the real output values. In each training round the ANN visits each training sample once. An example for the change of $A_1$ is
\begin{align}
A_1=\left(
\begin{array}{cc}
 0.34 & -0.87
\end{array}
\right)_{\text{init}}
&\xrightarrow{10~\text{rounds}}
A_1=\left(
\begin{array}{cc}
 1.15 & -0.93
\end{array}
\right)
\xrightarrow{20~\text{rounds}}
A_1=\left(
\begin{array}{cc}
 2.05 & -2.04
\end{array}
\right)\nonumber\\[3mm]
&\xrightarrow{30~\text{rounds}}
A_1=\left(
\begin{array}{cc}
 2.00 & -2.00
\end{array}
\right)
\end{align}
As we can see, already after 30 rounds of training, which takes about .3 seconds on a regular laptop, the network has learned, by looking at the examples in the training set, the relation between the input and the output value to a rather high accuracy.

\bigskip

While this was a rather trivial example for an application of type \ref{enum:reverse-engineer}, more complicated but still feasible examples are conceivable. One such example is the computation of line bundle cohomologies, which is also studied in Section \ref{sec:Example} as a simple application to evolving ANNs with genetic algorithms. In this case, the individual steps in the computation are known. Essentially one computes dimensions of ambient space cohomologies which are restricted to the Calabi-Yau hypersurface. Computing ambient space cohomology dimensions requires computing products and sums of the line bundle vectors. (To be more precise, depending on the sign of the first Chern class of the line bundle, the dimensions are either given in terms of binomial coefficients or are zero \cite{Anderson:2013qca}.) Restricting this to the Calabi-Yau involves twisting the line bundle with the normal bundle of the Calabi-Yau, which amounts to summing integers. Consequently, we expect that an ANN that contains layers that perform multiplications, additions, and set some input to zero can perform the exact computation needed in this task. The problem is that the number of additions and multiplications combined with setting dimensions of cohomology groups to zero becomes rather involved very quickly which prevents implementing the (in principle known) algorithm (it has been worked out for a simple complete intersection Calabi-Yau \cite{Buchbinder:2013dna}). In Section~\ref{sec:Example} we will evolve ANNs that solve the problem using regression (i.e.\ type \ref{enum:predict}) and defer the study of how efficiently an ANN can evolve into a complex function if provided with all building blocks involved in the computation (i.e.\ type \ref{enum:reverse-engineer}) to future research.

\bigskip
\noindent\textbf{Example: Classifying a line bundle as stable or unstable}\\\noindent 
In our next example, we look at a simple classification (i.e.\ type \ref{enum:classify}) problem. Suppose we want to build a heterotic string model. This requires specifying some Calabi-Yau threefold $X$ together with a vector bundle $\mathcal{V}$. For simplicity, we use a simple line bundle in this case, $\mathcal{V}=\mathcal{L}$. The line bundle has to satisfy a couple of constraints, one of which is $D$-flatness or bundle stability. A line bundle $\mathcal{L}$ satisfies the $D$-flatness criterion if 
\begin{align}
\label{eq:Stability}
\int_X J \wedge J \wedge c_1(\mathcal{L})=0 \,,
\end{align}
where $J=\sum a_i D_i$ is the K\"ahler form of the underlying Calabi-Yau $X$ and $c_1(\mathcal{L})=\sum k_i D_i$ is the first Chern class of the line bundle. Here, we have expanded the K\"ahler form in terms of a divisor basis $D_i$ of $X$ with the K\"ahler parameters $a_i$. Similarly, we have expanded $c_1(\mathcal{L})$ in terms of the first Chern class $k_i$ of the bundle on the $i^\text{th}$ divisor. Note that \eqref{eq:Stability} is generically only solved in a subspace of the full K\"ahler cone of $X$, i.e.\ it imposes a constraint on the K\"ahler parameters $a_i$. If the $D_i$ are chosen such that the K\"ahler cone is spanned by $a_i>0$ and the triple intersection numbers $\kappa_{ijk}=\int_X D_i D_j D_k$ are non-negative then a necessary condition for \eqref{eq:Stability} to have solutions is that some of the $k_i$ are positive and others negative. For concreteness, we study a complete intersection Calabi-Yau (the so-called ``bi-cubic'') which has two divisor classes that pull back from two ambient space factors ($\mathbbm{P}^2\times\mathbbm{P}^2$) and have intersection numbers
\begin{align}
\kappa_{111}=\kappa_{222}=0\,,\qquad \kappa_{112}=\kappa_{122}=3\,.
\end{align}

\begin{figure}[t]
  \centering
  \includegraphics[width=.5\textwidth]{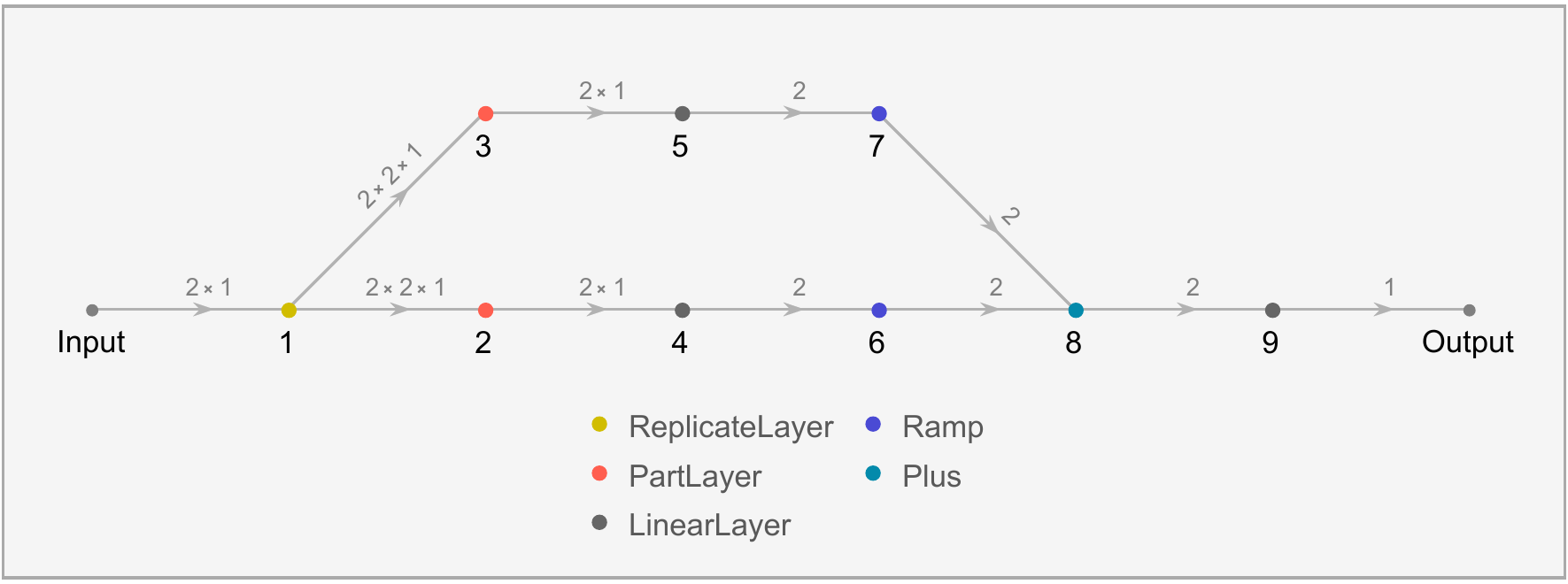}
  \caption{Layer representation of the ANN used to check bundle stability.}
  \label{fig:ANNStableBundle}
\end{figure}

\begin{figure}[t]
 \begin{tabular}{ccc}
  ~{\subfloat[][L6 Output port 1]{\includegraphics[width=.24\textwidth]{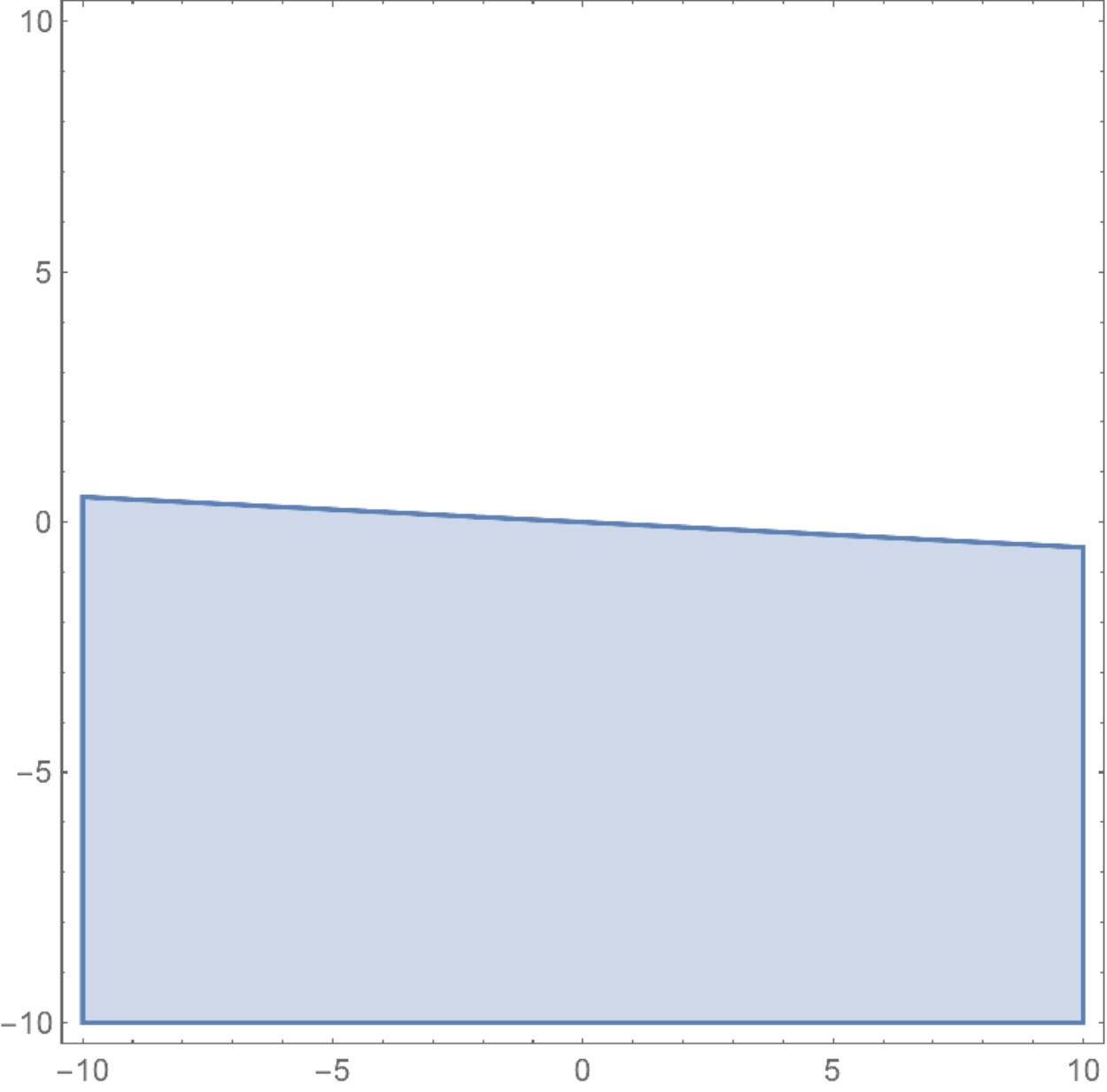}\label{fig:ANNStableRegionPlot1}}}&
  ~{\subfloat[][L6 Output port 2]{\includegraphics[width=.24\textwidth]{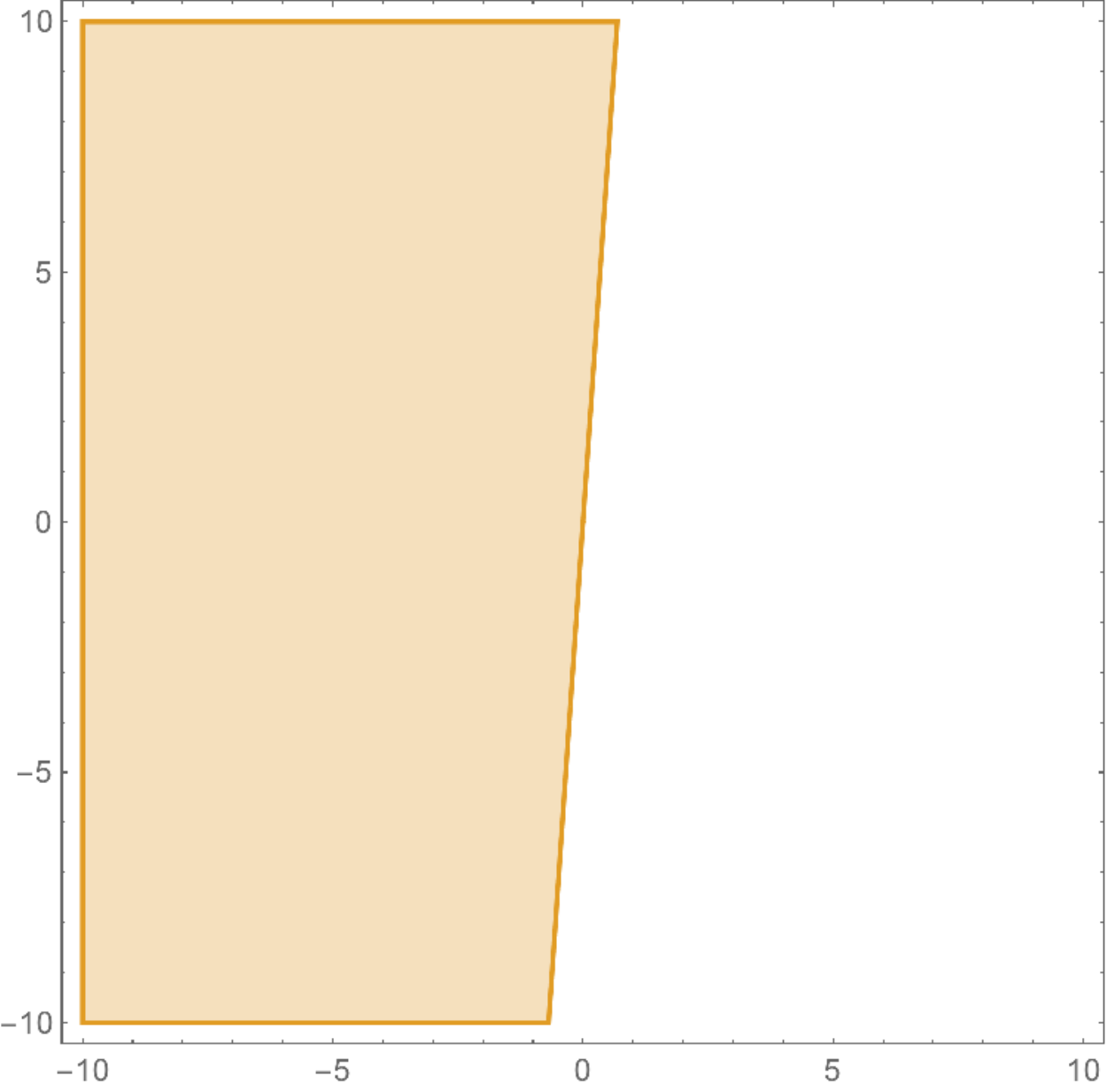}\label{fig:ANNStableRegionPlot2}}}&\\
  ~{\subfloat[][L7 Output port 1]{\includegraphics[width=.24\textwidth]{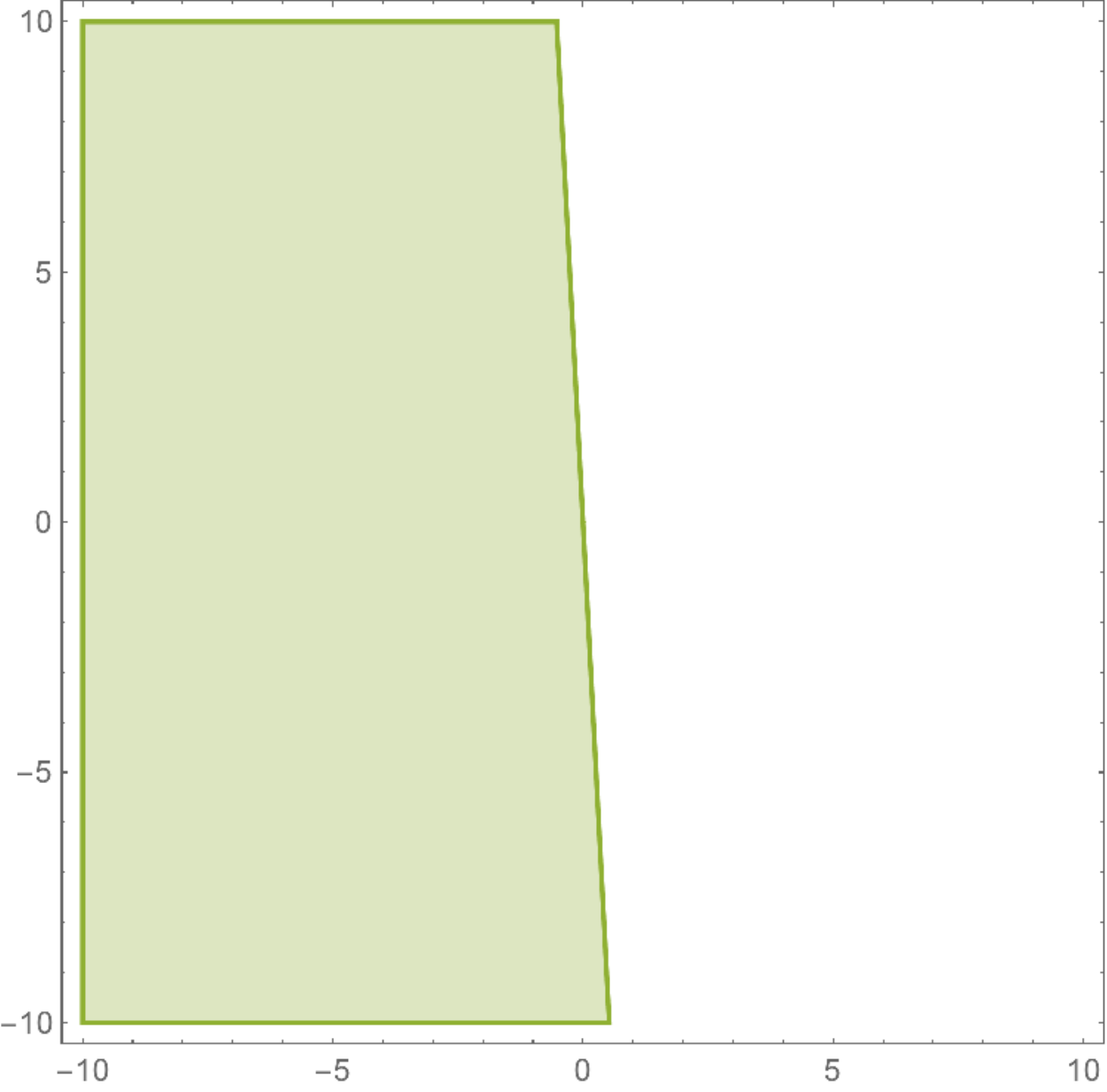}\label{fig:ANNStableRegionPlot3}}}&
  ~{\subfloat[][L7 Output port 2]{\includegraphics[width=.24\textwidth]{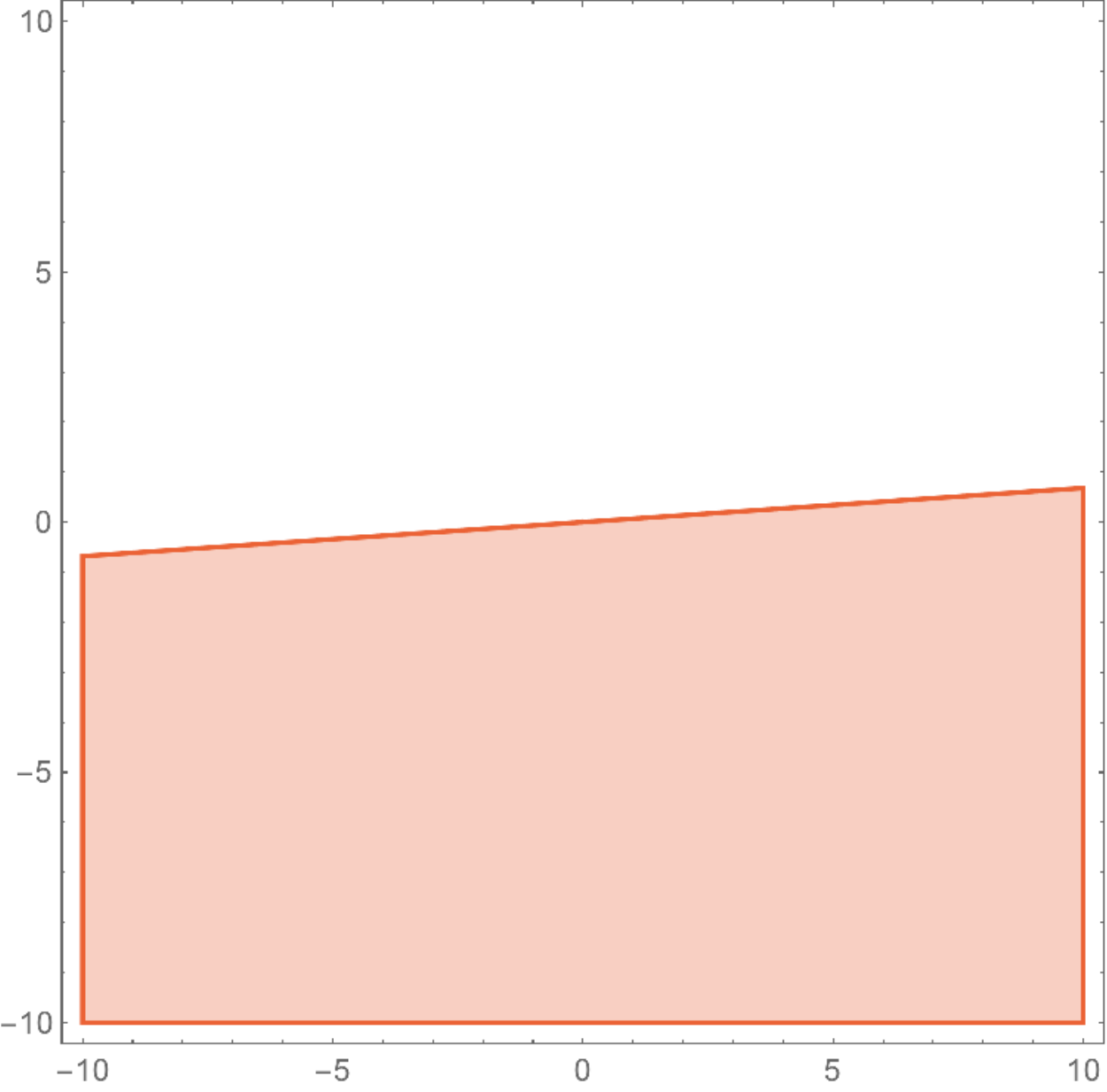}\label{fig:ANNStableRegionPlot4}}}&
  ~{\multirow[c]{-1}{*}[6.6cm]{\subfloat[][L9 Output port]{\includegraphics[width=.42\textwidth]{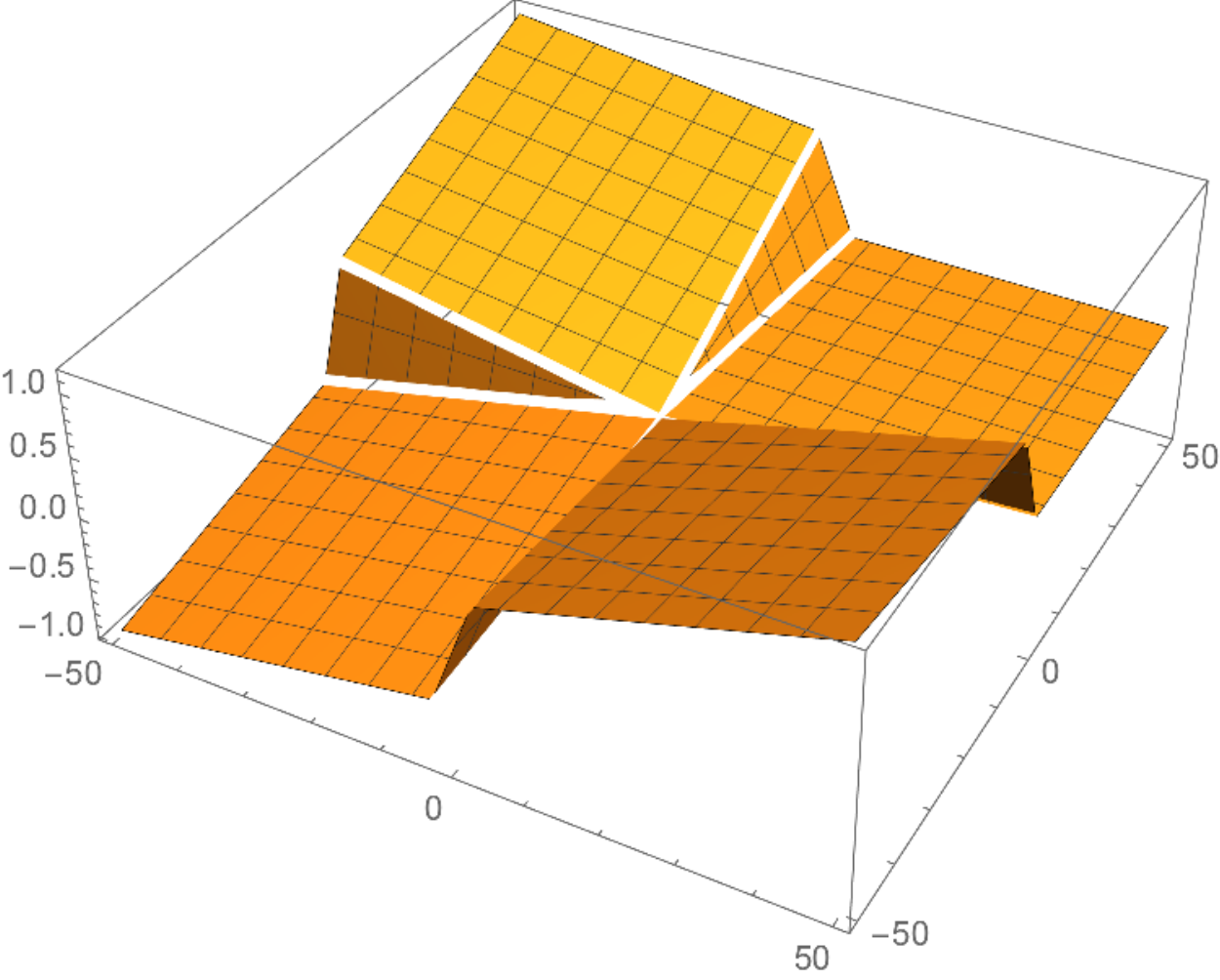}\label{fig:ANNStableValuesPlot}}}}\\[6mm]
  \end{tabular}
  \caption{The region plots \protect\subref{fig:ANNStableRegionPlot1} - \protect\subref{fig:ANNStableRegionPlot4} show the four output ports after the ramp layers L6 and L7. Plot \protect\subref{fig:ANNStableValuesPlot} shows the output after the last linear layer L9 on the vertical axis. All plots are in the $k_1$-$k_2$ plane.}
  \label{fig:ANNStablePlots}
\end{figure}

We now construct a random set of $10\,000$ line bundles with\footnote{Note that usually much smaller $k_i$ appear.} $k_i\in[-50,50]$ together with an integer indicating whether they are stable (1) or unstable (0). Out of the $10\,000$, we will train the network with a subset of $3\,000$ and check how well it performs on the other $7\,000$. The purpose of this introductory and pedagogical example is to illustrate the methods introduced above and to explain what the ANN does in a situation that is a bit more complex than the previous examples. We choose a network with a non-trivial topology and the number of nodes small enough such that we can easily visualize the different operations of the ANN in 3D plots. We therefore use an ANN which duplicates the input, processes it with a linear layer with two nodes, applies a ramp function, adds the two results obtained from the parallel process and maps them with another linear layer onto the output, which is a single number. The ANN is depicted in Figure~\ref{fig:ANNStableBundle}. After 15 seconds of training the network predicts the correct result with 90 percent accuracy. Training the ANN for a longer time does not improve this accuracy significantly. 

Let us investigate how the ANN predicts bundle stability and why further training does not help. The matrices $A_i$ that are applied in the linear layers 4, 5, and 9 of a trained network are given by
\begin{align}
A_4=\begin{pmatrix}0&-0.7\\-0.7&0\end{pmatrix}\,,\qquad A_5=\begin{pmatrix}-0.6&0\\0&-0.7\end{pmatrix}\,,\qquad A_9=\begin{pmatrix}-0.2&0.2\end{pmatrix}\,.
\end{align}
After layers 4 or 5, the input is passed on to the \textit{ramp layer}, which is a function layer that sets its input to zero if it is smaller than zero and acts as the identity if the input is larger than zero. This means that right before the entries are added element-wise in layer 8, the ramps have cut the input as shown in Figure \ref{fig:ANNStablePlots}\subref{fig:ANNStableRegionPlot1}-\subref{fig:ANNStableRegionPlot4}. Now, layer 8 adds the input of channels 1 and 3, and the ones of 2 and 4. After the addition, both entries contain everything but the first quadrant. In the next layer, i.e.\ layer 9, the two resulting numbers are multiplied by $A_9$ which subtracts the two entries. The result is shown as a 3D plot in the $k_1-k_2$ plane with the output value on the vertical axis in Figure \ref{fig:ANNStablePlots}\subref{fig:ANNStableValuesPlot}. 

While 90 percent accuracy is not bad there is the question why the neural network did not perform much better? The reason is that we used it as a predictor (i.e.\ all its layers map the input into $\mathbbm{R}$) even though it merely had to classify the input into two classes. The ANN tries to map the input as closely as possible to 0 if the bundle is unstable and to 1 if the bundle is stable. This also explains why the output behaves so strangely around the axes: On each axis, the bundle is unstable but on the next integral point away from the axis into the second or fourth quadrant the bundle is stable. Of course, the values 0 and 1 were assigned arbitrarily by us and do not have any physical significance. We can remedy this situation by introducing some normalization layer which normalizes the vector to the interval $[0,1]$. Then the ANN need not worry about reproducing the correct (arbitrary) numbers but only has to produce a number encoding the likelihood or probability of the outcome (i.e.\ ``stable'' or ``unstable''). Surely enough, after adding this normalization layer, and comparing its value (if it is larger than .5 the input is classified as stable and if it is less than or equal .5 the input is classified as unstable), the ANN reproduces the correct result in all cases after a few seconds of training. Alternatively, we could have used the one-hot encoding mentioned in the introduction and produced an output vector with two components, where the first encodes the likelihood of the input being stable and the second the likelihood of the input being unstable, and picked the most likely result. For this simple example with a binary outcome this is, however, unnecessary. 

This example also illustrates that an ANN that is expected to solve more complicated classification or prediction/regression tasks needs a more complicated structure. For example, the ANN we used first can only act by matrix multiplication and the identity (on top of cutting off negative values). If used in regression (on positive input values), this means that an ANN of this type can only perform linear regression, which is clearly a huge limitation and far from the power guaranteed for an ANN via the universal approximation theorem. Likewise, if we used an ANN with a single layer and a trivial or ramp activation function to classify data we could at best hope to classify data that is linearly separable. By building a deeper ANN or using more complicated functions we can again drastically improve its performance. If, on the other hand, the data would have been described best by linear regression, introducing non-linear layers is detrimental since it over-complicates the fitting function and can lead to strong oscillations or other problems.

\bigskip

Again, many other applications for classifications come to mind. These are particularly interesting since ANNs tend to perform better in classification problems as compared to regression problems. Besides the question of $D$-flatness, or bundle stability, studied here, one could ask many other phenomenology-inspired questions, such as
\begin{itemize}
  \item Does a given vector bundle lead to vector-like pairs of particles? We have investigated this question using an ANN which is trained on 3\,000 line bundle models for 90 seconds and used to classify 7\,000 other models according to whether or not they have vector-like pairs. The ANN correctly identified those models that have vector-like pairs in more than 85 percent of the cases.
  \item Does a given vector bundle satisfy the tadpole cancellation condition?
  \item Is a given model likely to have SUSY breaking, a small cosmological constant, lead to 60 e-folds of inflation,~\ldots?
\end{itemize}
Since especially the last questions are really hard to answer and it is a priori not clear what exactly determines these properties (although they can often be checked once one is given an explicit string model), it is a salient question how one should design ANNs that perform this classification task. We propose to use genetic algorithms to let the ANN ``design itself'' as we will explain next.

\section{Genetic algorithms}
\label{sec:GeneticAlgorithms}
The idea of genetic algorithms is to copy yet another mechanism of nature, namely evolution. The idea is to create a population of species, which would be a collection of ANNs in our case, and subject them to evolution. That means that they are allowed to mutate and reproduce, but only the fittest ones survive and are carried over to the next generation. 

Let us say we want to use an ANN as a predictor. If one knew the computational steps that are involved in computing the output from the input (as is the case when applying it to questions of type \ref{enum:reverse-engineer}) an ANN that uses these components will reproduce the correct output for any given input. However, the algorithm might be so involved that it is hard to actually implement it. In other cases (such as those arising in questions of type \ref{enum:predict}) one does not know the relation between the input and output data and we want to leave it to the ANN to figure out an efficient way of relating them. 

\begin{figure}
\centering
\includegraphics[width=.99\textwidth]{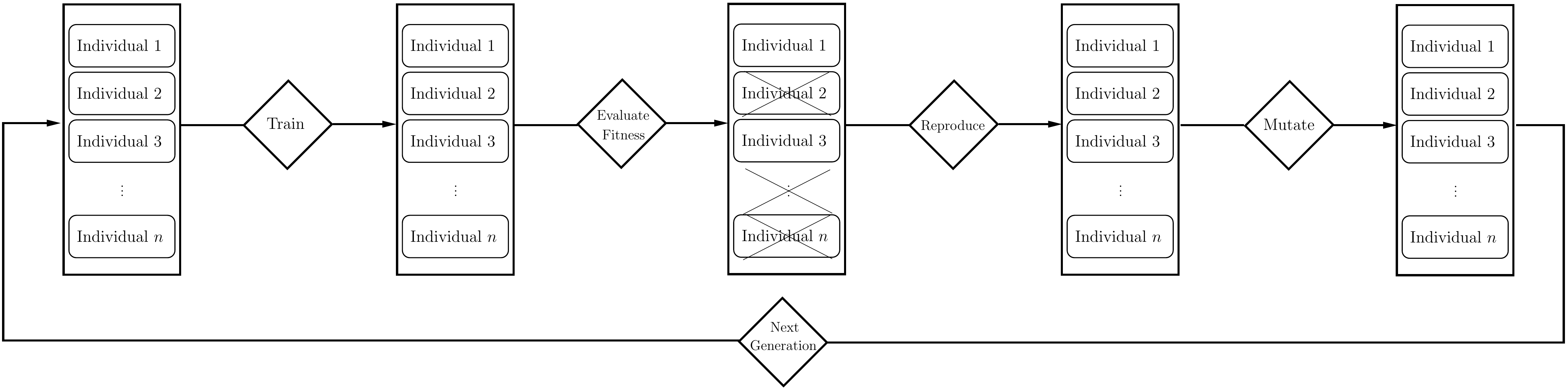}
\caption{Flowchart of the steps involved in the genetic algorithm: Starting from generation $i$, you train, evaluate the fitness, select the fittest one(s), let them reproduce, mutate, and start over.}
\label{fig:Evolution}
\end{figure}

This is where genetic algorithms are added to the picture. We start with a \textit{gene pool}\footnote{We chose to deviate from the notation used e.g.\ in \cite{Abel:2014xta} in which what we call the genome is called the genotype and what we call a gene is either a chromosome or an allele. The reason is that our gene pool contains elements of varying complexity; some are just single layers (i.e.\ alleles) while others are combinations thereof (i.e.\ chromosomes), and it would be tedious to make this distinction throughout.} consisting of a set of small functional ANN components. These might be either a single layer or a combination of layers that perform some action like multiplying two vectors component-wise. 

We then build a \textit{species}, consisting of a set of \textit{generations} of \textit{individuals} of ANNs (sometimes called \textit{population}). The individuals of the initial generation combine a random subset of the gene pool elements in some random order. Next, we train each individual (i.e.\ each ANN of the current generation) with our input data and check how well it performs on a set of unknown data. The one (or ones) that perform best are then carried over to the next generation. They then reproduce (either by cell division or by mating), passing on their \textit{genome} (i.e.\ their genes and the way in which they are combined) to the individuals of the next generation. 

During the reproduction phase, we allow for the genes to \textit{mutate} with a certain probability. During mutation, either new genes are added to the genome from the gene pool, existing genes are removed from the genome, or a gene of the genome is replaced by another gene from the gene pool. This is done by a process analogous to \textit{gene splicing} where we cut the ANN network between two genes and insert the new gene into the cut (we also allow this insertion to be at the beginning or end of the ANN). In order to ensure compatibility, we include appropriate connection layers into each gene such that they always expect a single input vector and produce a single output vector. In this way, the building blocks (genes) can be applied sequentially. In this simple approach, the non-trivial topology of the ANN is hidden within the genes, which are themselves put together in a sequential chain.

After the mutation phase, we start the whole process again with the new generation of individuals: train them on the data, evaluate their fitness, and let the fittest pass their genes on to the next generation. The process is summarized in Figure \ref{fig:Evolution}.

Note that there are several conceivable modifications to this approach which are, however, beyond the scope of the current paper:
\begin{itemize}
 \item Allow all individuals to reproduce, but give preference to the fitter ones (see e.g.\ in \cite{Abel:2014xta})
 \item During reproduction or mutation use larger parts of the genome rather than a single gene
 \item Allow changing the ``power'' of a given gene (e.g.\ its layer size) while keeping the actual structure intact
 \item Dynamically adjust the reproduction and mutation algorithm based on how complex the individual that has already evolved is
 \item Allow for more complex genes (e.g. with multiple inputs and outputs) to evolve more complicated ANN topologies (this requires a mechanism to ensure compatibility of the building blocks)
\end{itemize}

Using genetic algorithms for evolving neural networks rests on the assumption that individuals that have a subset of the genes that are needed to create the best-suited ANN in their genome are fitter (i.e.\ approximate the output better) than those that have genes that are unnecessary in the computation or at the wrong position of the computation sequence. There are several factors that govern whether the method will be successful. If e.g.\  the mutation rate is too low, one runs the risk of getting stuck in a ``local minimum''. This means that the initially fittest individual gets only slightly improved over the generations, but one never evolves a rather different individual which might perform much better. 

Similarly, if the training time is chosen inappropriately, one runs the risk of \textit{undertraining} or \textit{overtraining} the ANN. If the ANN is undertrained, it is not given enough time during training to find the best values for its trainable parameters, thus not realizing its full potential and eventually losing to another individual which has fewer trainable parameters and could thus find a better set of values for them during the short training time. If the ANN is overtrained, it has optimized its trainable parameters with respect to the training set. In overtraining, an individual with more trainable parameters will perform better simply because it can adjust more parameters to reproduce the output of the training set. However, the ANN might consist of a big number of complex layers, all of which have nothing to do with the computational steps that are actually needed to produce the output from the input, or it might have given too much emphasis to reproducing some features which are common to the models in the training set but not a common feature in general. In some sense, the ANN is \textit{overthinking} the solution to the problem. If e.g.\ handed the numbers $0,1,2,\ldots,6$ and asked for a prediction of the next number the answer could be 3 (if one was looking at the digits 26265647 to 26265653 of $\pi$) or 7 (if one is looking at a sequence of non-negative integers or applies Occam's razor).  Depending on the case, one needs a more complex ANN or a larger training set in order to distinguish them. Nevertheless, such overtrained ANN could out-perform another ANN that has evolved a subset of the correct layers. Once the species has evolved over many generations and brought forth a fit individual, one can look at it, check what it does to the input data, and in an ideal case reverse engineer the unknown function that maps the input to the output. However, this might be rather hard and really only useful in cases where the ANN has indeed evolved such that it reproduces the correct function, which might not be the case in many applications. 

Nevertheless, an ANN that predicts the outcome fairly well might still be preferable in many applications where the correct functions are known, e.g.\ if the ANN performs much faster. In some physical applications, it is in principle known how to compute a given quantity, but the computation can be rather costly in terms of computation time. For example, many applications in string model building requires solving equations over a ring (such as a polynomial ring or the ring of integers). Solving such problems involves computing Groebner Bases, solving Diophantic equations, computing resultants,\ldots, which are all very expensive steps. So, if one can just do these for a small subset and afterwards train an ANN to predict what the outcome for some given input data is without having to go through these computations every time, a huge performance improvement can be expected. In truly time-critical steps one could even factor the computation time into the fitness function.

While we have discussed the use of genetic algorithms for predictor ANNs, we can also use them to evolve an ANN that performs best in classifying a given set of data. The reasoning here is similar: If we either do not know exactly which computational steps are involved in generating the output or if it is not clear how to map the computation most efficiently onto an ANN, we can just start with a random set of ANNs and let them evolve into ANNs that perform much better in classifying the input. 

\section{Example: Evolving an ANN to compute bundle cohomology}
\label{sec:Example}

\subsection{The environment}
As an application we study the use of genetic algorithms to evolve an ANN that computes bundle cohomology of a line bundle $\mathcal{L}$ on a complete intersection Calabi-Yau threefold (number 6784 in the list of \cite{Candelas:1987kf}). This manifold has $h^{1,1}(X)=4$ and we thus need to specify the four first Chern classes of the line bundle on the four divisors. We produce a list of cohomology dimensions for line bundles whose first Chern classes are in the interval $[-3,3]$ using \cite{CICYPackage}. We divide this list of $7^4=2\,401$ line bundles into a training set with 600 elements and a validation set with $1\,800$ elements. This Calabi-Yau threefold is given in terms of three equations in a complex six-dimensional ambient space, i.e.\ it is codimension 3. The higher the codimension of a manifold, the costlier the computation of line bundle cohomologies becomes. While this codimension is still rather small, computing the list of bundle cohomologies took several hours.

\subsection{Setting up the various species}
In our setup, we are training 4 different species to compute the four cohomology dimensions  $h^i(\mathcal{L})$, $i=0,1,2,3$, independently. Of course, we could then just combine these 4 ANNs into a single ANN that computes all cohomology dimensions simultaneously. Or we could try to evolve an ANN that does just that by passing the entire vector $h^\bullet(\mathcal{L})$ to the genetic algorithm that evolves the ANN rather than each dimension individually.

\begin{figure}[H]
{
\centering
\begin{tabular}{|M{.3\textwidth}|M{.3\textwidth}|M{.3\textwidth}|}
\hline
\multicolumn{1}{|c|}{Linear Layer} & \multicolumn{1}{c|}{Ramp Layer} & \multicolumn{1}{c|}{Softmax Layer}\\
\includegraphics[width=.31\textwidth]{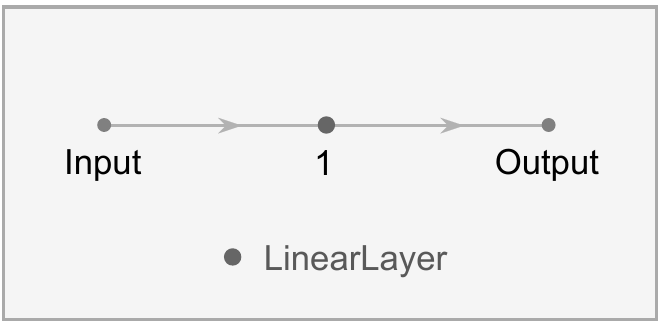}&\includegraphics[width=.31\textwidth]{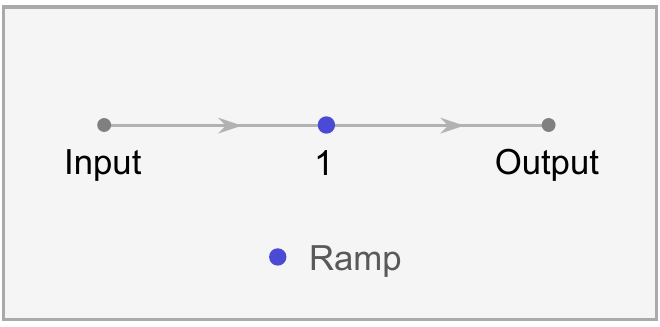}&\includegraphics[width=.31\textwidth]{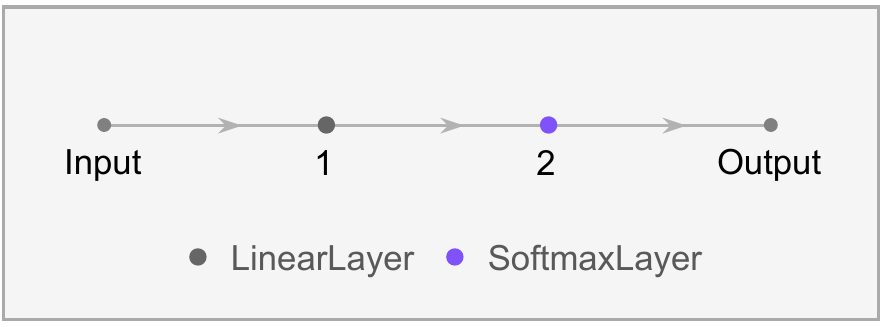}\\
Applies a linear transformation, $v_\text{o}=A\cdot v_\text{i}+b$ where $A$ is a matrix and $b$ is a vector.& Ramp activation function, which sets negative values to zero and acts as the identity on positive values.& Applies an exponential normalization function to the input.\\
\hline
\hline
\multicolumn{1}{|c|}{Tanh Layer} &  \multicolumn{1}{c|}{Logistic Sigmoid Layer} & \multicolumn{1}{|c|}{LSTM Layer}\\
\includegraphics[width=.31\textwidth]{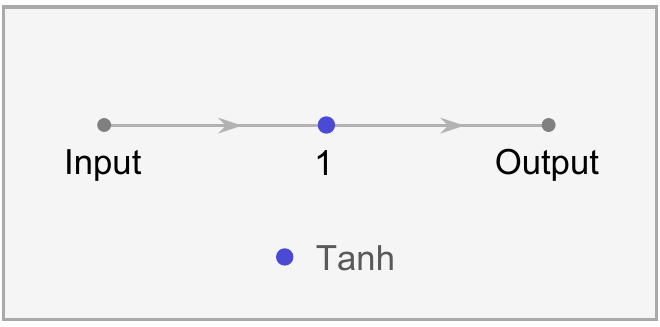}&\includegraphics[width=.31\textwidth]{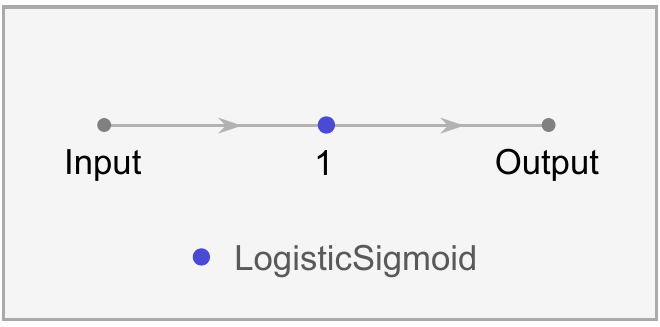}&\includegraphics[width=.31\textwidth]{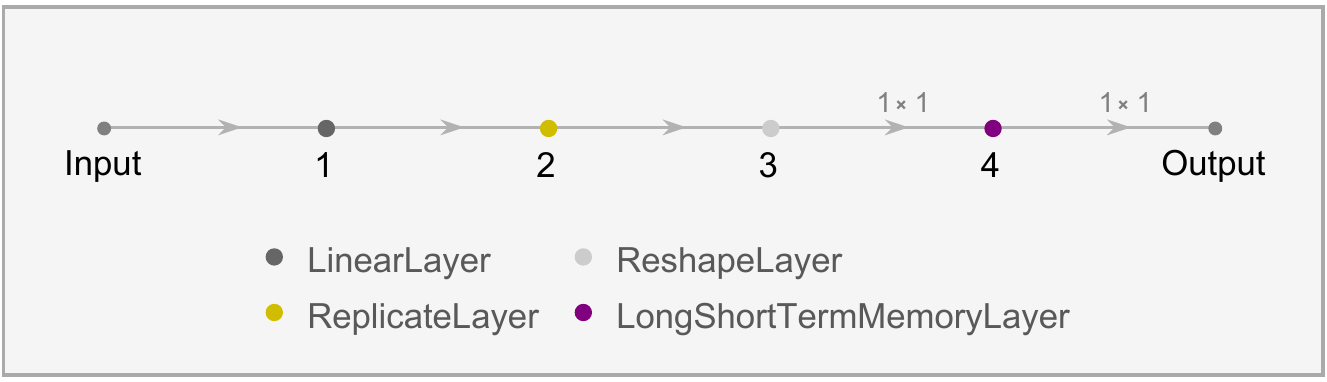}\\
Hyperbolic tangent activation function, which maps its input into the interval $[-1,1]$. & Logistic sigmoid activation function $1/(1+e^{-x})$. This is similar to the Tanh layer, but it maps the input into the interval $[0,1]$.& The Long Short-Term~Me\-mory layer is a more complex layer using feed-back. This gene includes some connection layers to prepare the input for the LSTM layer.\\
\hline
\hline
 \multicolumn{1}{|c|}{Plus Layer}& \multicolumn{1}{c|}{Times Layer} &\\
\includegraphics[width=.31\textwidth]{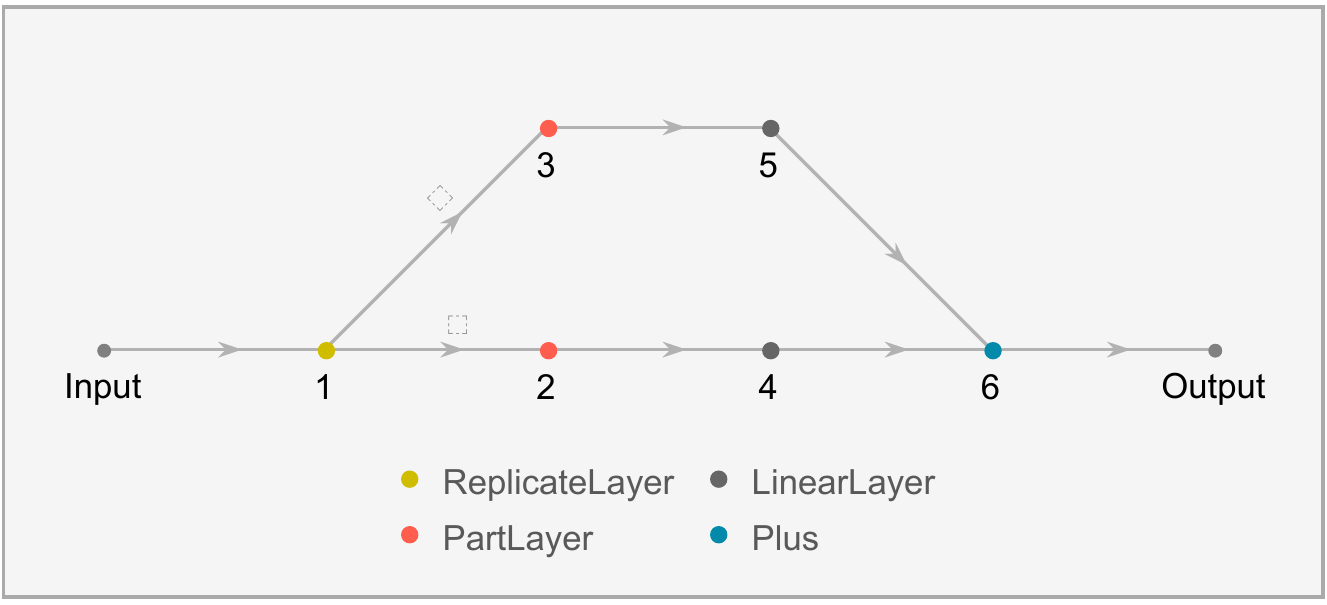}&\includegraphics[width=.31\textwidth]{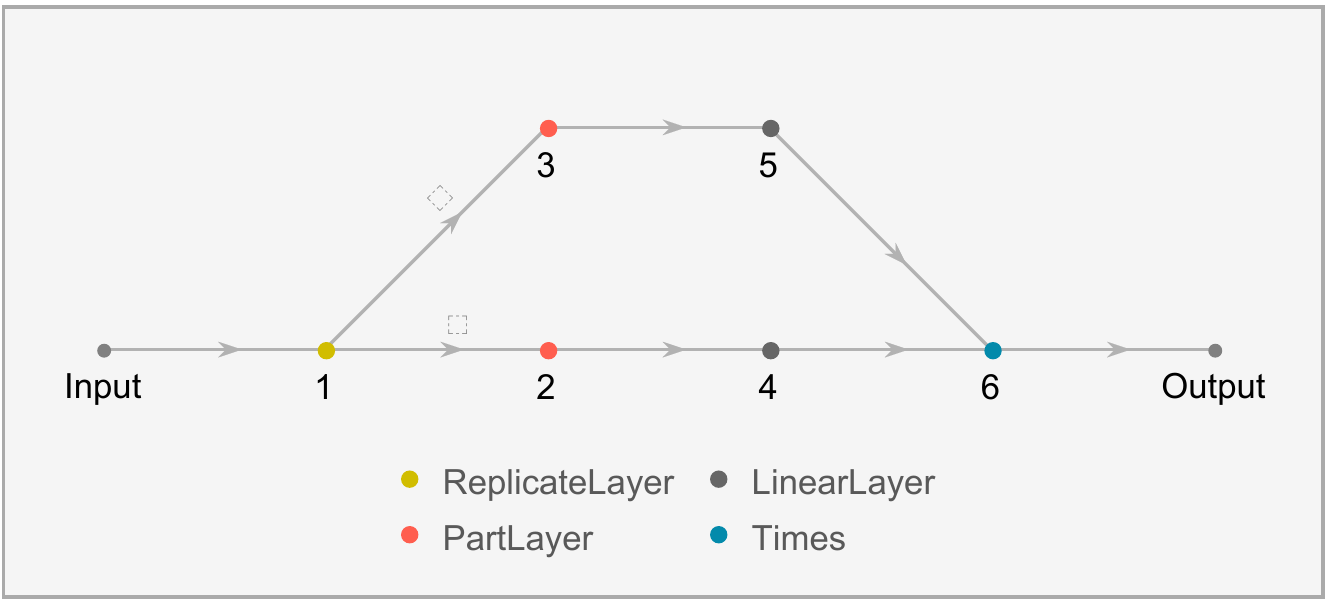}&\\
This layer adds its input vector component-wise. Connection layers prepare the input for this binary operation.&This layer multiplies its input vector component-wise. Connection layers prepare the input for this binary operation.&\\
\hline
\end{tabular}
}
\caption{Overview of the gene pool used in the evolution of the neural network that computes line bundle cohomologies. The ANNs are given in their layer representation.}
\label{fig:GenePool}
\end{figure}

The first generation of each species is created with up to four randomly selected elements from the gene pool, cf.\ Figure \ref{fig:GenePool}. In each generation, the two fittest individuals survive and produce seven children. In our investigation, we have tested reproduction via cell division, mating, and a combination of the two. By comparing results and by following the lineage of the fittest individual of each generation we found that for not very complex species it is actually better to reproduce via cell division. Note that this is similar to the evolution process in nature: Also there more complex species use more complex reproduction mechanisms while simpler organisms from which 
the complex species evolved are reproducing via cell division. So in this case we let each of the two fittest individuals reproduce via cell division (i.e.\ clone themselves) with a probability of 0.5. 
%
\linebreak\noindent
\noindent The mutation rate is set to 10 percent, we train each individual for 60 seconds, stop after 10 generations and allow for a maximum of 400 perceptrons per layer.

\subsection{Analysis of the result}
We find that the fittest ANNs correctly predict $h^0(\mathcal{L})$ and $h^3(\mathcal{L})$ in 82 percent of the cases while $h^1(\mathcal{L})$ and $h^2(\mathcal{L})$ are predicted correctly in 73 percent of the cases. Once evolved and trained, computing the bundle cohomology dimensions with the ANNs takes 7 seconds on a current laptop, which is a drastic improvement compared to the several hours it takes to compute the bundle cohomology dimensions exactly using Koszul and Leray spectral sequences.

\begin{figure}[t]
  \centering
  \subfloat[][Change of maximal fitness.]{\includegraphics[width=.465\textwidth]{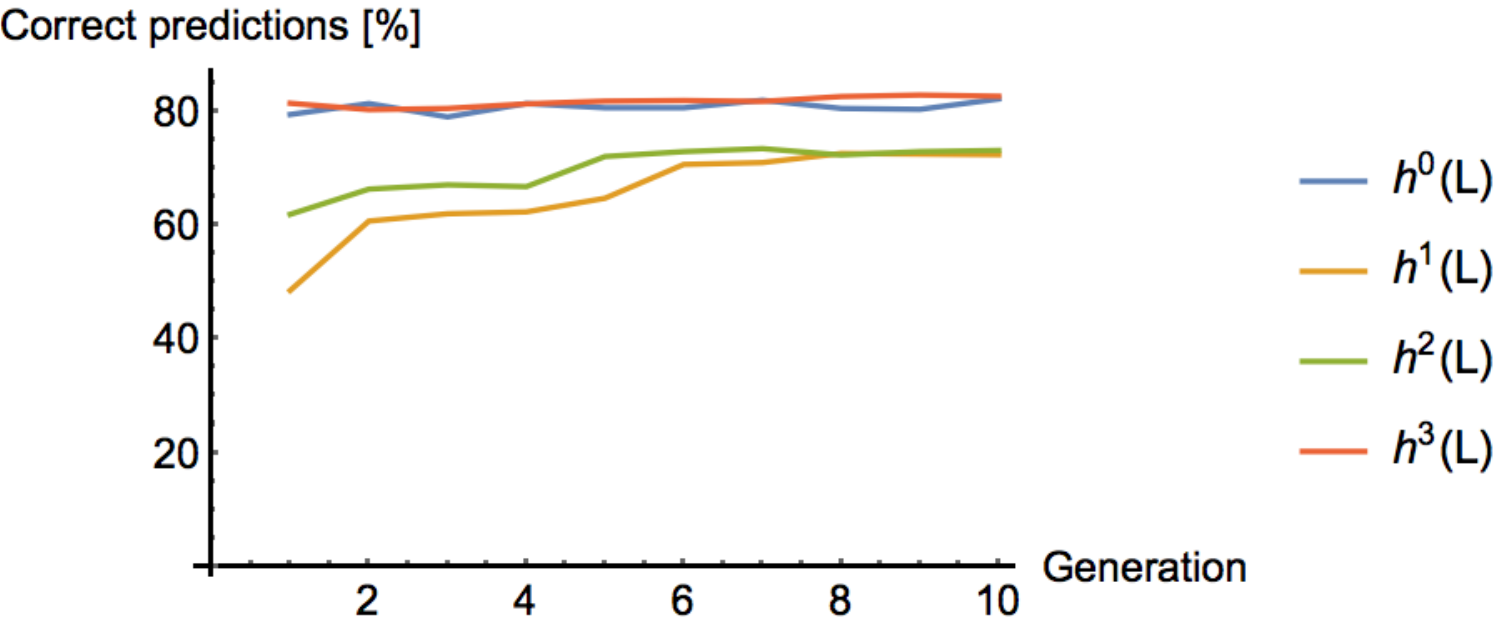}\label{fig:MaxFitness}}~~~~
  \subfloat[][Change of average fitness.]{\includegraphics[width=.465\textwidth]{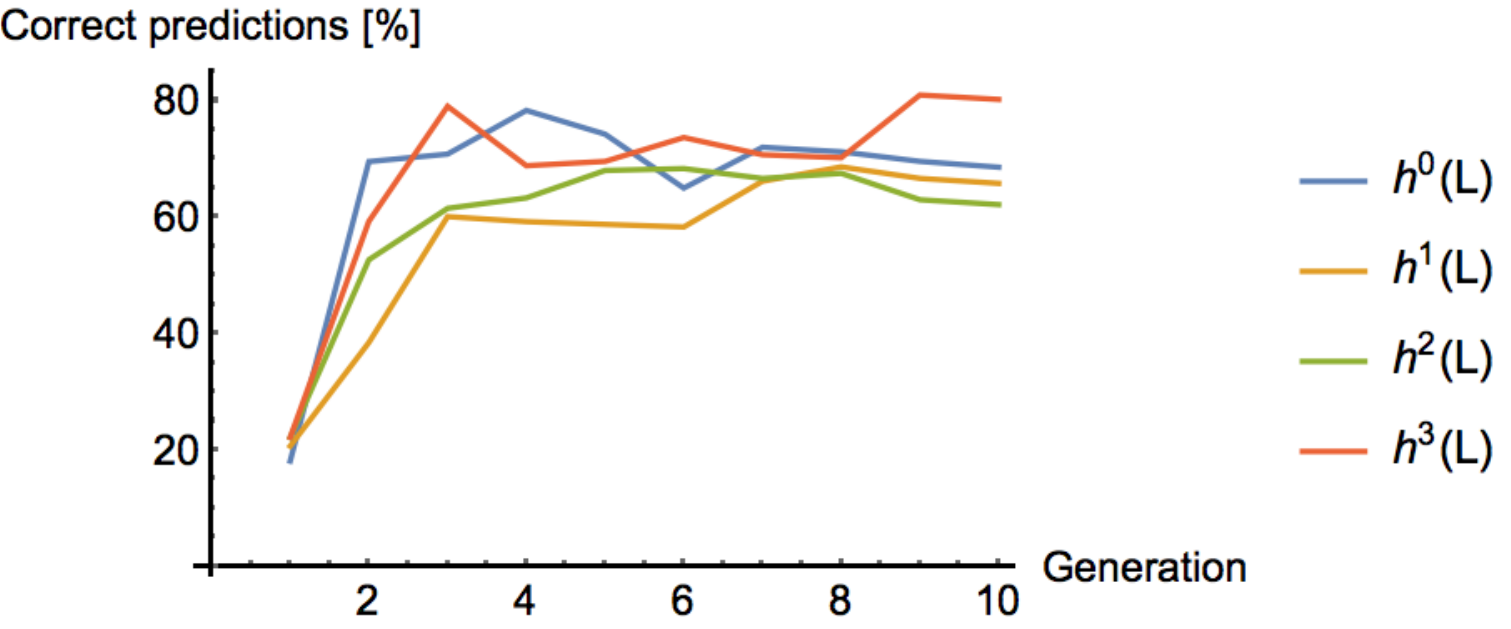}\label{fig:AvgFitness}}
  \caption{Change of \protect\subref{fig:MaxFitness} maximal and \protect\subref{fig:AvgFitness} average fitness over 10 generations for the four species predicting $h^i(\mathcal{L})$.}
  \label{fig:EvolutionResultsExample}
\end{figure}

The difference in performance between $h^0,h^3$ vs $h^1,h^2$ stems most likely from the fact that the cohomology groups of the former tend to be zero more often\footnote{Note that bundle stability, which is a rather mild constraint for this simple setup, implies $h^0=0$, since otherwise the trivial bundle can inject into the vector bundle and destabilize it. Furthermore, by looking at the maps in the Koszul resolution or at how the Leray spectral sequence stabilizes in the cohomology computations, the projective ambient space cohomology dimensions as computed from the Bott-Borel-Weil and the K\"unneth formula in combination with the Kodaira vanishing theorem (which applies due to the range of the first Chern classes of the line bundle and normal bundle of the Calabi-Yau), it can be seen that $h^0$ receives fewer contributions than e.g.\ $h^1$. Via Serre duality, similar results apply to $h^3$.} while the latter are more often non-trivial. Furthermore, in computing these numbers many different intermediate cohomology dimensions have to be computed which are then combined using a Koszul or a Leray spectral sequence. In both cases the middle cohomologies tend to receive more non-trivial contributions. We have summarized the evolution of the fittest individual of each species over the 10 generations in Figure \ref{fig:EvolutionResultsExample}\subref{fig:MaxFitness}. Figure \ref{fig:EvolutionResultsExample}\subref{fig:AvgFitness} shows the change in average fitness over the generations. As we can see, after six or seven generations both the maximal and average fitness is not increased significantly anymore. This could mean that the neural network has at this point evolved the fittest species that is possible with the gene pool provided. This raises the interesting question of how to design or choose a gene pool such that a sufficiently complex ANN can be built from its genes in the most efficient way (from the universal approximation theorem we learn that in theory we only need a few layers, but these need to contain a huge number of nodes, which most likely makes this solution very unfeasible). It is also possible that we are stuck in a local maximum at this point. In order to get beyond this point, one could try to adapt the parameters governing the ANN dynamically, such as the reproduction mechanism, the mutation rate, the number of children, the training time, $\ldots$ While these interesting questions are beyond the scope of the current investigation we hope to revisit them in the future.

We observe that the two ANNs for predicting $h^0(\mathcal{L})$ and $h^3(\mathcal{L})$ have evolved independently into a similar species (i.e.\ there genome is similar), as have the ANNs for predicting $h^{1}(\mathcal{L})$ and $h^2(\mathcal{L})$. In particular, both $h^1$ and $h^2$ have evolved an LSTM layer, hinting at their more complex nature. Note that, by virtue of Serre duality and the symmetric choice $[-3,3]$ as an interval for the first Chern classes, the training and validation sets of $h^0$ and $h^3$, as well as for $h^1$ and $h^2$, are physically equivalent. Let us also remark that training the ANNs longer does not improve their performance; after 60 seconds of training they have already learned everything they can from the training set they are provided.

\subsection{Applicability to other environments}
As explained above, cohomology computations become increasingly unfeasible in terms of computation time with growing codimension of the manifold. In contrast, the performance time of the ANN does not change with the codimension, since it simply applies a number of functions to its input values. In order to check how the ANNs that have evolved for computing line bundle cohomologies for a specific Calabi-Yau perform on another we reuse them and train them with new input data of the new Calabi-Yau. To be precise, we have used the ANNs that evolved for computing cohomologies on the complete intersection Calabi-Yau 6784 to compute cohomologies on the complete intersection Calabi-Yau 7862.  The latter is a codimension 1 rather than a codimension 3 hypersurface in a different projective ambient space, but it also has four divisor classes, which means that the input format (i.e.\ the first Chern classes of the line bundle) is the same. The result shows that they perform equally well in this different environment if they are trained with the data for this environment. 

In the future, we plan on extending the setup to provide both the data that specifies the vector bundle and the Calabi-Yau. In such setups, many questions arise. First and foremost, it would be very interesting to check whether we can train such an ANN with a subset of bundles and Calabi-Yaus and extract either predictions or classifications for other models on other Calabi-Yaus. If this turns out to be feasible one can train the network on data which is computationally more accessible in order to study models that are at the boundary or even beyond the current computational power of modern PCs. In addition, these ANNs could be used to study different but related questions. For example, if one has evolved an ANN that can compute vector bundle cohomology, one could just give it the tangent bundle as a vector bundle and compute the Hodge numbers of the Calabi-Yau manifold.

\section*{Acknowledgments}
I would like to thank Sven Krippendorf and Andre Lukas for their encouraging feedback. My work is supported by the EPSRC grant EP/N007158/1 ``Geometry for String Model Building''.


\begingroup\begin{thebibliography}{10}

\bibitem{Candelas:1987kf}
P.~Candelas, A.~M. Dale, C.~A. Lutken, and R.~Schimmrigk ``{Complete
  Intersection Calabi-Yau Manifolds},'' {\em Nucl. Phys.} {\bf B298} (1988)
493.

\bibitem{Kreuzer:2000xy}
M.~Kreuzer and H.~Skarke ``{Complete classification of reflexive polyhedra in
  four-dimensions},'' {\em Adv. Theor. Math. Phys.} {\bf 4} (2002) 1209--1230
\href{http://www.arXiv.org/abs/hep-th/0002240}{[{\tt hep-th/0002240}]}.

\bibitem{Morrison:2012js}
D.~R. Morrison and W.~Taylor ``{Toric bases for 6D F-theory models},'' {\em
  Fortsch. Phys.} {\bf 60} (2012) 1187--1216
\href{http://www.arXiv.org/abs/1204.0283}{[{\tt 1204.0283}]}.

\bibitem{Morrison:2012np}
D.~R. Morrison and W.~Taylor ``{Classifying bases for 6D F-theory models},''
  {\em Central Eur. J. Phys.} {\bf 10} (2012) 1072--1088
\href{http://www.arXiv.org/abs/1201.1943}{[{\tt 1201.1943}]}.

\bibitem{Gray:2013mja}
J.~Gray, A.~S. Haupt, and A.~Lukas ``{All Complete Intersection Calabi-Yau
  Four-Folds},'' {\em JHEP} {\bf 07} (2013) 070
\href{http://www.arXiv.org/abs/1303.1832}{[{\tt 1303.1832}]}.

\bibitem{Anderson:2015iia}
L.~B. Anderson, F.~Apruzzi, X.~Gao, J.~Gray, and S.-J. Lee ``{A new
  construction of Calabi–Yau manifolds: Generalized CICYs},'' {\em Nucl.
  Phys.} {\bf B906} (2016) 441--496
\href{http://www.arXiv.org/abs/1507.03235}{[{\tt 1507.03235}]}.

\bibitem{Davey:2009bp}
J.~Davey, A.~Hanany, and J.~Pasukonis ``{On the Classification of Brane
  Tilings},'' {\em JHEP} {\bf 01} (2010) 078
\href{http://www.arXiv.org/abs/0909.2868}{[{\tt 0909.2868}]}.

\bibitem{Anderson:2013xka}
L.~B. Anderson, A.~Constantin, J.~Gray, A.~Lukas, and E.~Palti ``{A
  Comprehensive Scan for Heterotic SU(5) GUT models},'' {\em JHEP} {\bf 01}
  (2014) 047
\href{http://www.arXiv.org/abs/1307.4787}{[{\tt 1307.4787}]}.

\bibitem{Nilles:2014owa}
H.~P. Nilles and P.~K.~S. Vaudrevange ``{Geography of Fields in Extra
  Dimensions: String Theory Lessons for Particle Physics},'' {\em Mod. Phys.
  Lett.} {\bf A30} (2015) no.~10, 1530008
\href{http://www.arXiv.org/abs/1403.1597}{[{\tt 1403.1597}]}.

\bibitem{Nibbelink:2015ixa}
S.~Groot~Nibbelink, O.~Loukas, F.~Ruehle, and P.~K.~S. Vaudrevange ``{Infinite
  number of MSSMs from heterotic line bundles?},'' {\em Phys. Rev.} {\bf D92}
  (2015) no.~4, 046002
\href{http://www.arXiv.org/abs/1506.00879}{[{\tt 1506.00879}]}.

\bibitem{Blaszczyk:2015zta}
M.~Blaszczyk, S.~Groot~Nibbelink, O.~Loukas, and F.~Ruehle ``{Calabi-Yau
  compactifications of non-supersymmetric heterotic string theory},'' {\em
  JHEP} {\bf 10} (2015) 166
\href{http://www.arXiv.org/abs/1507.06147}{[{\tt 1507.06147}]}.

\bibitem{Franco:2016qxh}
S.~Franco, S.~Lee, R.-K. Seong, and C.~Vafa ``{Brane Brick Models in the
  Mirror},'' {\em JHEP} {\bf 02} (2017) 106
\href{http://www.arXiv.org/abs/1609.01723}{[{\tt 1609.01723}]}.

\bibitem{Halverson:2016tve}
J.~Halverson and J.~Tian ``{Cost of seven-brane gauge symmetry in a quadrillion
  F-theory compactifications},'' {\em Phys. Rev.} {\bf D95} (2017) no.~2,
  026005
\href{http://www.arXiv.org/abs/1610.08864}{[{\tt 1610.08864}]}.

\bibitem{Halverson:2017ffz}
J.~Halverson, C.~Long, and B.~Sung ``{On Algorithmic Universality in F-theory
  Compactifications},''
\href{http://www.arXiv.org/abs/1706.02299}{[{\tt 1706.02299}]}.

\bibitem{Douglas:2003um}
M.~R. Douglas ``{The Statistics of string / M theory vacua},'' {\em JHEP} {\bf
  05} (2003) 046
\href{http://www.arXiv.org/abs/hep-th/0303194}{[{\tt hep-th/0303194}]}.

\bibitem{Kachru:2003aw}
S.~Kachru, R.~Kallosh, A.~D. Linde, and S.~P. Trivedi ``{De Sitter vacua in
  string theory},'' {\em Phys. Rev.} {\bf D68} (2003) 046005
\href{http://www.arXiv.org/abs/hep-th/0301240}{[{\tt hep-th/0301240}]}.

\bibitem{He:2017aed}
Y.-H. He ``{Deep-Learning the Landscape},''
\href{http://www.arXiv.org/abs/1706.02714}{[{\tt 1706.02714}]}.

\bibitem{Krefl:2017yox}
D.~Krefl and R.-K. Seong ``{Machine Learning of Calabi-Yau Volumes},''
\href{http://www.arXiv.org/abs/1706.03346}{[{\tt 1706.03346}]}.

\bibitem{Allanach:2004my}
B.~C. Allanach, D.~Grellscheid, and F.~Quevedo ``{Genetic algorithms and
  experimental discrimination of SUSY models},'' {\em JHEP} {\bf 07} (2004) 069
\href{http://www.arXiv.org/abs/hep-ph/0406277}{[{\tt hep-ph/0406277}]}.

\bibitem{Abel:2014xta}
S.~Abel and J.~Rizos ``{Genetic Algorithms and the Search for Viable String
  Vacua},'' {\em JHEP} {\bf 08} (2014) 010
\href{http://www.arXiv.org/abs/1404.7359}{[{\tt 1404.7359}]}.

\bibitem{Cybenko:1989aa}
G.~Cybenko ``{Approximations by superpositions of sigmoidal functions},'' {\em
  Mathematics of Control, Signals, and Systems} {\bf 2} (1989) 303.

\bibitem{Anderson:2013qca}
L.~B. Anderson, J.~Gray, A.~Lukas, and B.~Ovrut ``{Vacuum Varieties,
  Holomorphic Bundles and Complex Structure Stabilization in Heterotic
  Theories},'' {\em JHEP} {\bf 07} (2013) 017
\href{http://www.arXiv.org/abs/1304.2704}{[{\tt 1304.2704}]}.

\bibitem{Buchbinder:2013dna}
E.~I. Buchbinder, A.~Constantin, and A.~Lukas ``{The Moduli Space of Heterotic
  Line Bundle Models: a Case Study for the Tetra-Quadric},'' {\em JHEP} {\bf
  03} (2014) 025
\href{http://www.arXiv.org/abs/1311.1941}{[{\tt 1311.1941}]}.

\bibitem{CICYPackage}
L.~B. Anderson, J.~Gray, S.-J. Lee, Y.-H. He, and A.~Lukas ``{A 2009 ‘CICY
  package’, based on methods described in
  \href{http://www.arXiv.org/abs/0911.1569}{[{\tt 0911.1569}]},
  \href{http://www.arXiv.org/abs/0911.0865}{[{\tt 0911.0865}]},
  \href{http://www.arXiv.org/abs/0805.2875}{[{\tt 0805.2875}]},
  \href{http://www.arXiv.org/abs/hep-th/0703249}{[{\tt hep-th/0703249}]},
  \href{http://www.arXiv.org/abs/hep-th/07022109}{[{\tt hep-th/0702210}]}}.''

\end{thebibliography}\endgroup
\providecommand{\href}[2]{#2}
\end{document}